\documentclass[aps,prb,twocolumn,groupedaddress,amsmath,amssymb,showpacs]{revtex4}

\usepackage{color}

\usepackage{graphicx}
\usepackage{hyperref}
\usepackage{pifont}

\begin{document}
\title{\textit{Ab-initio} study on the possible doping strategies for MoS$_2$ monolayers}
\author{Kapildeb Dolui, Ivan Rungger, Chaitanya Das Pemmaraju, and Stefano Sanvito}
\affiliation{School of Physics and CRANN, Trinity College, Dublin 2, Ireland}
\date{\today}
\begin{abstract}
Density functional theory is used to systematically study the electronic and magnetic properties of doped MoS$_2$ monolayers, where
the dopants are incorporated both via S/Mo substitution or as adsorbates. Among the possible substitutional dopants at the Mo site, Nb 
is identified as suitable p-type dopant, while Re is the donor with the lowest activation energy. When dopants are simply adsorbed on a
monolayer we find that alkali metals shift the Fermi energy into the MoS$_2$ conduction band, making the system n-type. Finally, the 
adsorption of charged molecules is considered, mimicking an ionic liquid environment. We find that molecules adsorption can lead to 
both n- and p-type conductivity, depending on the charge polarity of the adsorbed species.
\end{abstract}
\maketitle
\section{Introduction}

In recent years two dimensional (2D) materials have attracted a growing interest due to their potential for future 
nano-electronics applications, owing to their unusual physical, optical and electrical properties arising from the quantum 
confinement associated to their ultra thin structure~\cite{RPP_2011_74}. These considerations apply particularly well to 
layered transition metal di-chalcogenides (LTMDs) in which a multitude of electronic states have been observed, namely 
metallicity, semi-conductivity and charge density waves. Given the richness in the electronic properties it is not a surprise 
that this materials class has been called out as the ideal platform for a multitude of applications~\cite{Adv_1969_18, Adv_1987_36}. 
For example, the prototypical LTMD, molybdenum di-sulfide MoS$_2$, has been widely explored as lubricant~\cite{Wear_1967_10}, 
catalyst~\cite{CatT_2005_107} and as lithium ion battery anode~\cite{ChAJ_2012_7}. 

Structurally MoS$_2$ consists of covalently bonded S-Mo-S 2D hexagonal planes (monolayers), which in the bulk are bound together 
in a layered structure by weak van der Waals forces~\cite{JPCM_2000_12}. Interestingly, the electronic properties of bulk MoS$_2$ 
show a strong dependence on the layer thickness~\cite{NL_2010_10_1271}. Single MoS$_2$ monolayers are particularly intriguing.
These display a direct band-gap of 1.9~eV, which makes them a semiconducting alternative to gapless graphene. Indeed the band-gap 
of graphene can be opened by fabricating nanoribbons~\cite{PRL_2008_100_206803} or by depositing it over a suitable 
substrate~\cite{Nature_2007_6_770}. However this comes at the cost of deteriorating the carrier mobility due to edge and impurity 
scattering~\cite{PNAS_2007_104_18392}. In contrast, the absence of dangling bonds, the high crystallinity, and the low dimensionality, 
make the performance of LTMDs comparable to those of currently existing Si transistors at the scaling 
limit~\cite{JAP_2007_101, IEEE_2011_58, NL_2012_12_1538}. 
MoS$_2$ monolayer based transistors have been recently demonstrated to operate at room temperature, with a mobility of 
at least 200 cm$^{2}$/Vs and on/off current ratios of 10$^8$ with low standby power dissipation~\cite{Nature_2011_6}. Interestingly, both 
n-type~\cite{Nature_2011_6, PRL_2010_105, Adv_2012_24_2320, Small_2012_8_682} and 
p-type~\cite{ACIEd_2011_50_11093,Small_2012_8_966} conductivities have been reported for ultra-thin MoS$_2$ layers, depending 
on the experimental details. The origin of such diverse conducting properties remains to date far from being clear.

The possible creation of Mo and/or S vacancies in MoS$_2$ monolayers during the growth cannot be used as a doping strategy, 
since those vacancies always produce deep trap states in the MoS$_2$ monolayer band-gap~\cite{PRL_2004_92}. In a recent study we have investigated how
unintentional doping of the substrate holding a MoS$_2$ monolayer may determine its conductivity~\cite{arXiv:1301.2491}. 
Here we explore possible strategies for doping directly into the MoS$_2$ monolayer. Although there are studies available on specific 
dopants for single layer MoS$_2$ \cite{PRL_2004_92, APL_2010_96_082504, JPChC_2011_115_13303}, 
to our knowledge there is not yet a systematic investigation comparing the doping properties of the various defects, when they are
incorporated in either a substitutional position or they are adsorbed on the surface. Moreover, it has been reported that in a liquid-gated 
electric double layer transistor, the ions can modulate the electronic properties from insulating to metallic through electrostatic induced 
carrier doping in the transport channel at a finite gate bias~\cite{NL_2012_12_2988}. Compared with other dopants, such as metals and 
light atoms, ionic liquids have several advantages as surface dopants: a large variety of such ionic complexes are available, they can 
accumulate more carriers than conventional solid state gated transistors, and they do not induce structural disorder like substitutional 
dopants. In fact, ambipolar transistor operation has been achieved using thin flakes of MoS$_2$ in an ionic liquid gated 
environment~\cite{NL_2012_12_1136}. The present work explores also this possibility.

Here we systematically study the effects of dopants on the electronic structure of a MoS$_2$ monolayer, by calculating the formation 
energies and the electronic properties of halogens, non-metals, transition metals and alkali metals added in various geometrical configurations. 
Our aim is that of identifying potential candidates for making MoS$_2$ monolayers either n- or p-type in a controlled way. Then we present 
an investigation of the electronic structure of MoS$_2$ monolayers when two molecular ions, NH$_4^+$ and BF$_4^-$, are adsorbed at 
the surface. The paper is organized as follows. After a brief discussion of the computational methods used we turn our attention to the case
of substitutional doping by looking at both at the S and the Mo site. Next we consider the case of doping by adsorption of alkali and molecular
ions. At the end of the results section we briefly discuss the robustness of our calculations against the particular choice of exchange and 
correlation functional and finally we conclude.

\section{Methodology}
\begin{figure}
\center
\includegraphics[width=8.0cm,clip=true]{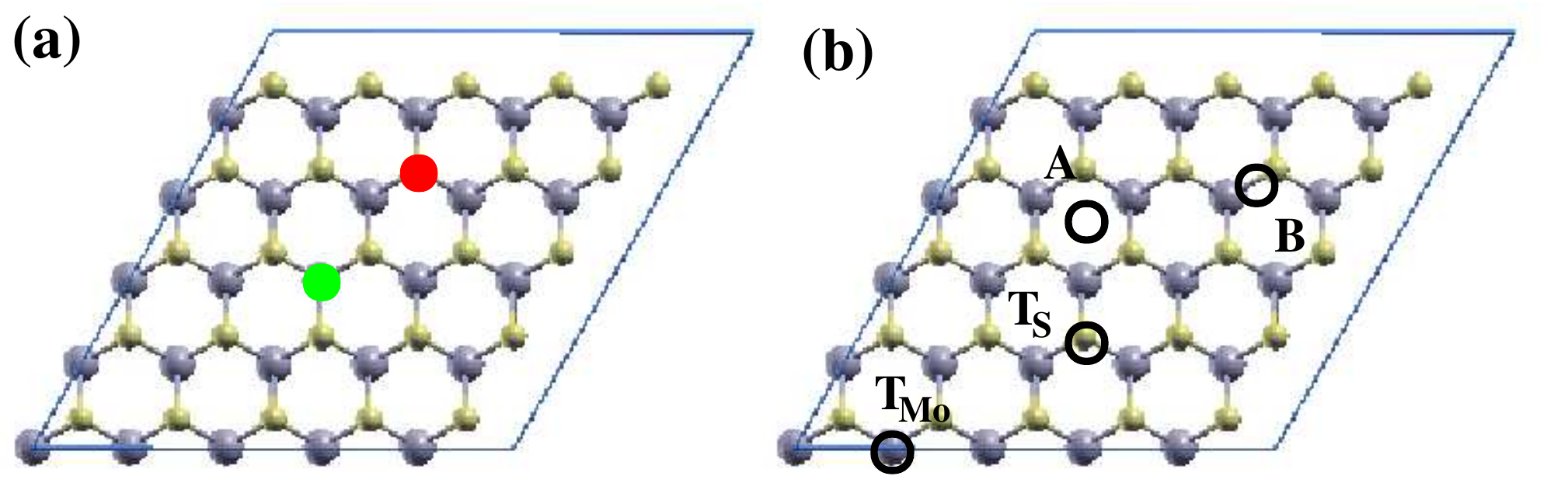}
\caption{(Color online) Structure of the supercell used for the calculations. In (a) we show a 5$\times$5 MoS$_2$ monolayer supercell 
with the possible substitutional dopants. Mo is substituted by a transition metal (green dot): Y, Zr, Nb, Re, Rh, Ru, Pd, Ag, Cd; S is substituted 
by nonmetals and halogens (red dot): P, N, As; F, Cl, Br, I. In (b) we display the same supercell where we indicate the possible adsorption sites 
defined in the text. Light gray spheres indicate Mo atoms, light yellow spheres indicate S.}
\label{fig:ImpurityCell}
\end{figure}

In order to investigate the electronic properties of a MoS$_2$ monolayer doped with impurities, \textit{ab-initio} calculations are performed 
using density functional theory~\cite{PRB_1964_136, PRA_1965_140} within the local spin density approximation (DFT-LSDA) for the 
exchange and correlation potential. In particular we consider the Ceperly-Alder LSDA parametrization~\cite{PRL_1980_45} as implemented 
in the SIESTA code~\cite{IOP_2002_14}. In our calculations double-$\zeta$ polarized~\cite{PRB_2001_64} numerical atomic orbitals basis 
sets are used for all atoms, and the Troullier-Martins scheme is used for constructing norm-conserving pseudopotentials~\cite{PRB_1991_43}. 
A 5$\times$5 hexagonal supercell [see Fig.~\ref{fig:ImpurityCell}(a)] with (15.66$\times$15.66)~{\AA}$^2$ lateral dimensions is constructed
and doping is introduced by replacing/adding a single atom in the supercell. This corresponds to simulating a Mo$_{1-x}$A$_x$S$_{2-y}$B$_y$
periodic system with $x=4\%$ ($y=2\%$) when the doping is at the Mo (S) site. An equivalent plane wave cutoff of 250~Ry is used for the real 
space mesh and the Brillouin zone is sampled over a 5$\times$5$\times$1 Monkhorst-Pack $k$-grid. Periodic boundary conditions are applied and a 
vacuum layer of at least 15 {\AA} is placed above the monolayer to minimize the interaction between the adjacent periodic images. A temperature 
of 300 K is used when populating the electronic states with a Fermi distribution. The relaxed geometries are obtained by conjugate gradient, 
where all the atoms in the supercell are allowed to relax until any force is smaller than 0.02 eV/\AA.

In order to verify that the calculated impurity level alignments are robust against the choice of exchange correlation functional, we repeat 
the calculations for the main results by using the screened hybrid functional of Heyd-Scuseria-Ernzerhof (HSE06) \cite{HSE}. All DFT 
calculations based on the HSE06 functional are carried out with the projector augmented wave (PAW) pseudo potential plane-wave 
method\cite{PRB_1994_50_17953} as implemented in the VASP code \cite{PRB_1996_54_11169}. A 3$\times$3$\times$1 
Monkhorst-Pack\cite{PRB_1976_13_5188} $k$-point grid is employed and the plane wave energy cutoff is 500~eV.

The formation energy of a particular substitutional dopant, $E_\mathrm{form}$, is defined as
\begin{eqnarray}
E_\mathrm{form} = E_\mathrm{tot}\mathrm{(MoS_2^*+X)}+E_\mathrm{bulk}\mathrm{(host)}\nonumber \\-E_\mathrm{tot}\mathrm{(MoS_2)} - E_\mathrm{bulk}\mathrm{(X)}\:,
\end{eqnarray}
where $E_\mathrm{tot}$(MoS$_2^*$+X) is the total energy of the system including the substitutional atom X, $E_\mathrm{tot}$(MoS$_2$) 
is the total energy of the corresponding pristine MoS$_2$ monolayer, while $E_\mathrm{bulk}$(X) and $E_\mathrm{bulk}$(host) are 
respectively the energies of the substitutional atom X and of the substituted Mo (S) host atom in their bulk (diatomic molecule) forms. 
In contrast the formation energy for an adsorbate, $E_\mathrm{ads}$, can be written as
\begin{eqnarray}
E_\mathrm{ads} = E_\mathrm{tot}\mathrm{(MoS_2+Y)}-E_\mathrm{tot}\mathrm{(MoS_2)}-E_\mathrm{bulk}\mathrm{(Y)}\:,
\end{eqnarray}
where $E_\mathrm{tot}$(MoS$_2$+Y) refers to the total energy when the adsorbate Y is attached to MoS$_2$, and $E_\mathrm{bulk}$(Y) is the energy of an 
adsorbate Y in its bulk form. In order to find the most stable configuration for adsorption, we consider four possible positions labeled as 
follows: T$_\mathrm{S}$ (adsorbate on top of S), T$_\mathrm{Mo}$ (adsorbate on top of Mo), A (adsorbate above the center of the hexagonal 
ring of MoS$_2$), and B (adsorbate above the middle of the Mo-S bond) [see Fig.~\ref{fig:ImpurityCell}(b)].

%
\section{Results and discussion}
\begin{figure}
\center
\includegraphics[width=8.0cm,clip=true]{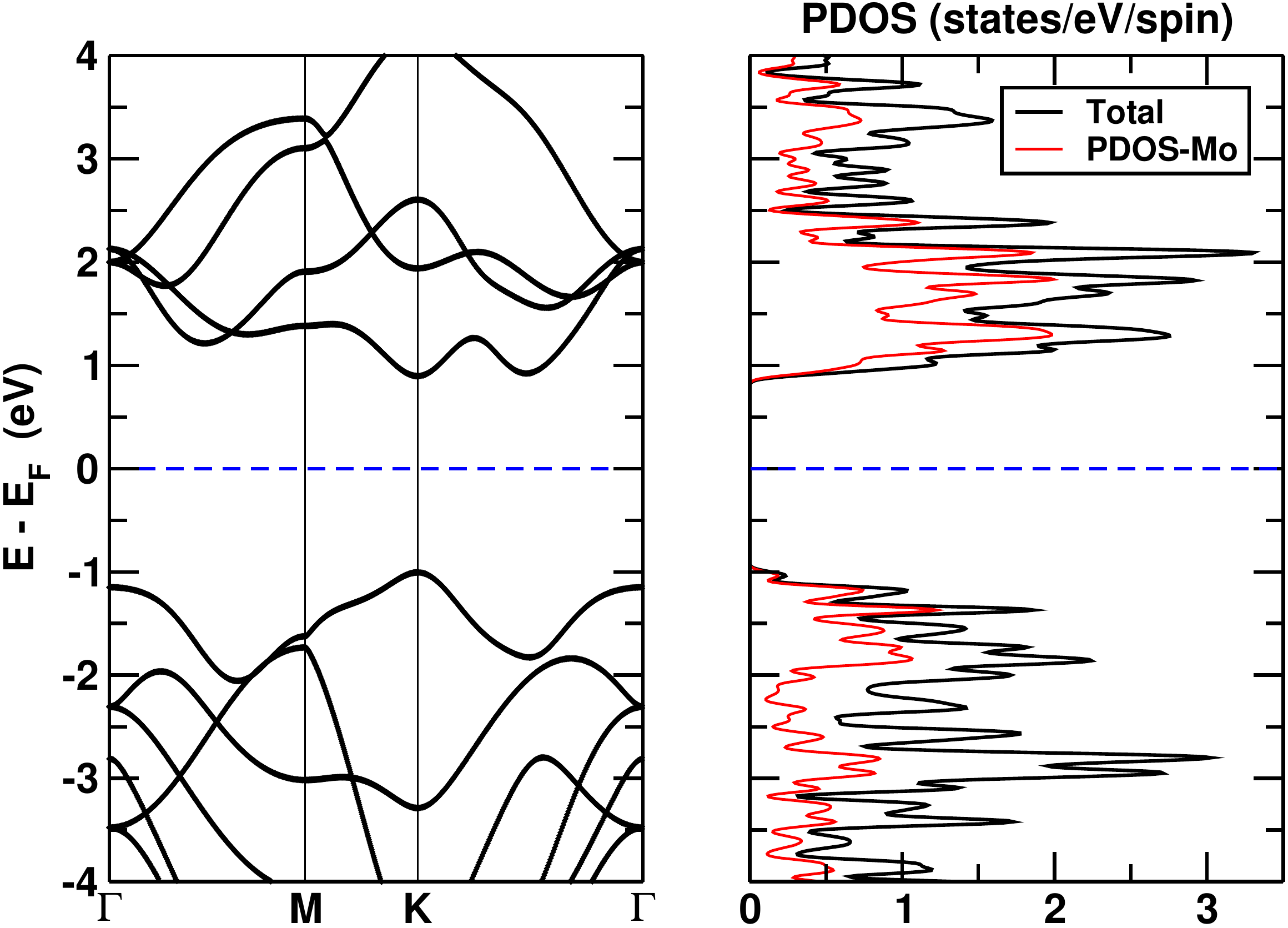}
\caption{(Color online) Band-structure (left) and DOS (right) of a pristine (undoped) MoS$_2$ monolayer. The red curve indicates the 
DOS projected on the Mo atoms.}
\label{fig:1L_band_PDOS}
\end{figure}

Before discussing the formation energy and the electronic structure of the possible dopants let us here briefly review the electronic
properties of a single MoS$_2$ monolayer. As the MoS$_2$ thickness is decreased from bulk to a few layers, the valence band 
minimum shifts from half way along $\Gamma$-K line towards K ~\cite{NL_2010_10_1271}. For a monolayer 
the band-gap becomes direct at K (see Fig.~\ref{fig:1L_band_PDOS}), a transition which has been recently observed 
experimentally~\cite{NL_2010_10_1271}. The computed LSDA band-gap of 1.86~eV is in good agreement with the experimental 
optical band-gap of 1.90 eV~\cite{PRL_2010_105}, although such an agreement has to be considered fortuitous. In fact the 
absorption edge of an optical excitation measures the energy difference between the quasi-particle band-gap and the exciton binding 
energy. In MoS$_2$ monolayers this latter is of the order of 1~eV, as confirmed recently by many-body calculations~\cite{Ramasubramaniam}. 
Thus, one expects that the true quasi-particle spectrum has a band-gap of approximately 2.9~eV, in good agreement with that computed 
with the $GW$ scheme, either at the first order level~\cite{Ramasubramaniam} (2.82~eV) or self-consistently~\cite{Lamb} (2.76~eV). 
At the end of this section we will discuss how band-gap corrections, via the HSE06 functional, affect our main results. 
In any case, for both bulk and monolayer the band-structure around the band-gap is derived mainly from Mo-4$d$ orbitals (see 
the orbital projected density of state, PDOS, in Fig.~\ref{fig:1L_band_PDOS}), although there are also small contributions in the valence 
band from the S-3$p$ ones. 

\begin{table}
\begin{tabular}{ccc}\hline
Impurity& $m$($\mu_B$/impurity) & E$_\mathrm{form}$(eV)\\
\hline\hline F, Cl, Br &  1.00 & 0.60 (F), 1.93 (Cl), 2.16 (Br) \\ 
I&0.00& 2.53 \\ \hline
N, P&  1.00&2.90 (N), 1.89 (P)\\
As&0.00&1.91\\ \hline
Re&1.00& 2.05\\
Ru&2.00 &3.05\\
Rh&3.00&4.15\\
Pd&4.00& 5.63\\
Ag& 2.57& 7.28\\
Cd&1.58& 6.11\\ \hline
Nb, Zr, Y&0.00& -0.19 (Nb), -0.48 (Zr), 0.52 (Y)\\ 
\hline
\end{tabular}
\caption{The theoretically calculated magnetic moment, $m$, and the formation energy, $E_\mathrm{form}$ of different substitutional 
dopants in the MoS$_2$ monolayer.}
\label{formation_0}
\end{table}

\subsection{Substitutional doping}\label{substitutional}
\subsubsection*{Substitution at the S site}
\begin{figure}
\center
\includegraphics[width=8.0cm,clip=true]{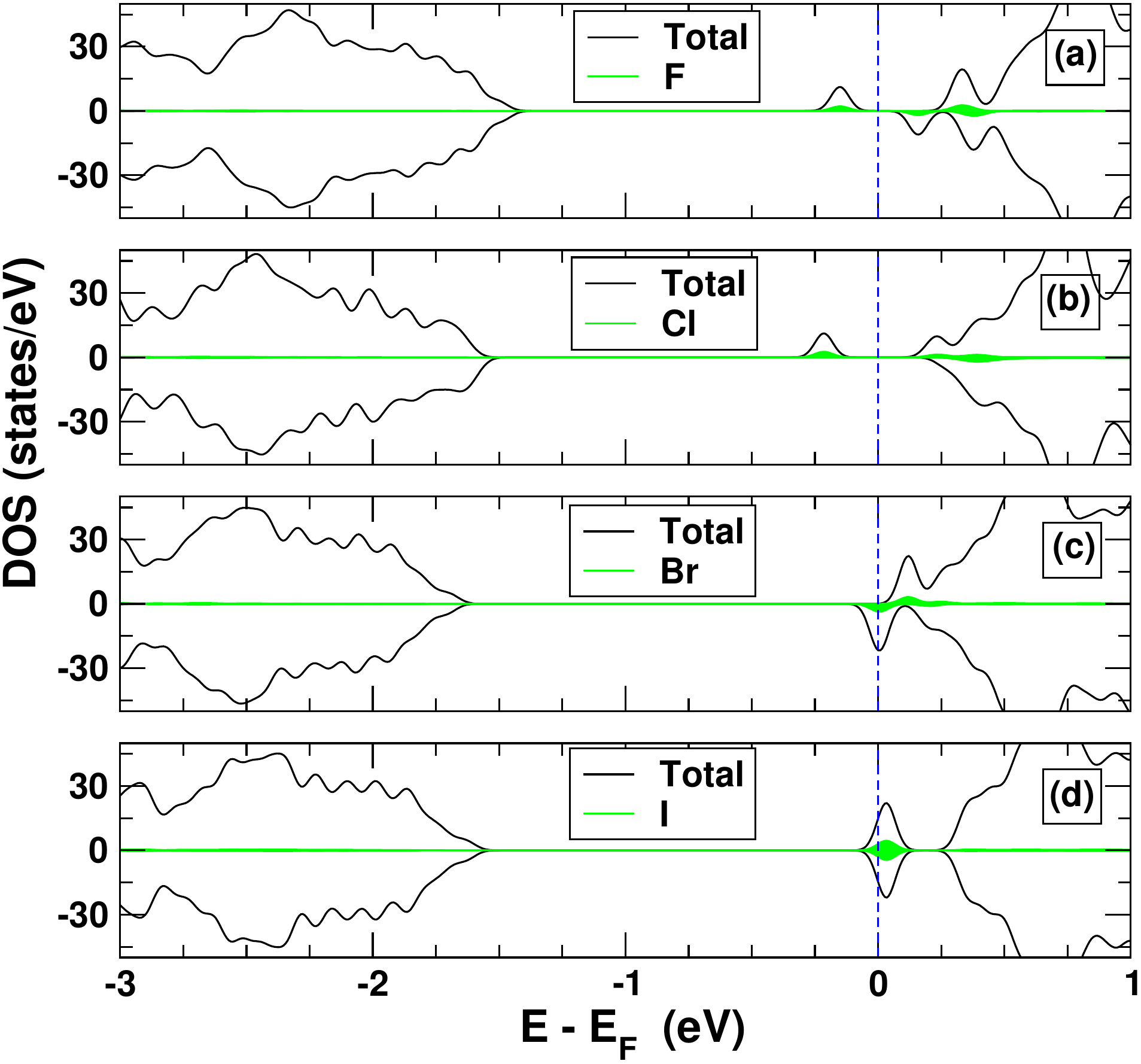}
\caption{(Color online) DOSs for a 5$\times$5 MoS$_2$ supercell in which one S atom is replaced by (a) F, (b) Cl, (c), Br and (d) I. 
Negative DOS values refer to minority spins, while positive are for the majority. The blue dashed line indicates the Fermi energy, 
while the colored shaded areas indicate the DOS projected over the dopants. Note that all the DOSs are aligned to have a common 
Fermi level, $E_\mathrm{F}=0$.}
\label{fig:PDOS_FClBrI}
\end{figure}

We begin our analysis by substituting a surface S atom with elements taken from the halogen family, namely F, Cl, Br and I. These are
expected to act as a source of n-type doping for MoS$_2$, since they have one additional $p$ electron with respect to S. The DOSs for
the supercell including one atom of the halogen family are presented in Fig.~\ref{fig:PDOS_FClBrI}. As a representative system, we discuss 
in details results for Cl-doping, since the other halogens present a similar electronic structure. A Cl-doped MoS$_2$ monolayer has a 
magnetic ground state, where an occupied defect level is formed at about 0.4~eV below the conduction band minimum (CBM). 
The corresponding minority state is located above the CBM and is empty. These defect levels originate from the hybridization between 
the Cl-3$p$ and the Mo-4$d$ states. Similarly to the Cl substitution, for both F and Br dopants (isoelectronic to Cl) the system is 
para-magnetic, having a magnetic moment of approximately 1~$\mathrm \mu_B$. In the case of F substituting S, the spin-splitting is 
larger than that of Cl, whereas it is smaller for the Br substitutional case. Eventually, when S is replaced by iodine, the impurity becomes 
nonmagnetic, although the impurity level is still located well below (0.3 eV) the CBM [see Fig.~\ref{fig:PDOS_FClBrI}(d)]. 
Note that having an impurity presenting a ground state with finite magnetic moment has nothing to do with diluted ferromagnetism in 
$p$-type systems, as sometimes erroneously claimed in literature. This simply indicates that the additional electron remains unpaired 
and localized around the impurity \cite{Andrea}. The dependence of the spin-splitting of the impurity level on the atomic number 
simply follows the well-known trend for the exchange-correlation integral~\cite{Janak}.

\begin{figure}
\center
\includegraphics[width=8.0cm,clip=true]{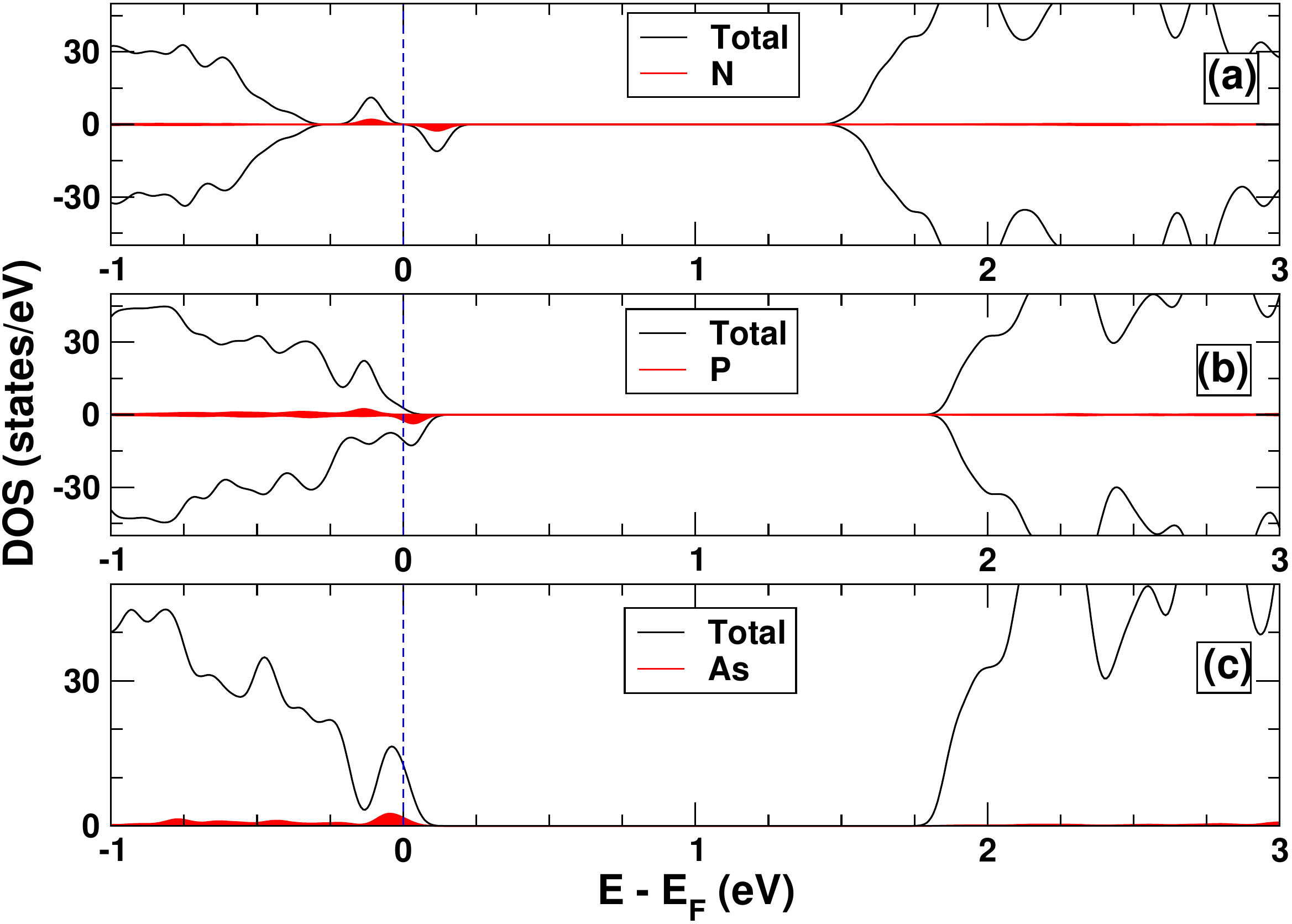}
\caption{(Color online) DOSs for a 5$\times$5 MoS$_2$ supercell in which one S atom is replaced by (a) N, (b), P and (c) As. 
Negative values refer to minority spins, positive values to majority spins. The blue dashed line indicates the Fermi energy, the colored 
shaded areas indicate the DOS projected on the dopants. Note that all the DOSs are aligned to have a common Fermi level, 
$E_\mathrm{F}=0$.}
\label{fig:PDOS_NPAs}
\end{figure}

We then investigate the possibility of p-type doping obtained by replacing a S atom with a group V element of the periodic table, 
namely N, P and As (see Fig.~\ref{fig:PDOS_NPAs}). In the case of N the supercell has a magnetic ground state [see 
Fig.~\ref{fig:PDOS_NPAs}(a)], with a magnetic moment of 1$\mu_\mathrm{B}$ and a spin-splitting of about 0.20~eV. In both 
the spin channels gap states are introduced above the valence band maximum (VBM), and these are formed by hybridization between 
the N-2$p$ and the Mo-4$d$ valence orbitals. Similarly, when substituting S with P, the system is again magnetic, but the defect states 
are rigidly shifted closer to the VBM of the pristine MoS$_2$ monolayer in both the spin channels. The spin-splitting remains similar to that of
N. Finally for the As substitutional case, the defect states shift even more towards the VBM. Additionally, the spin-splitting vanishes and 
the system becomes a p-type semiconductor. The partially unoccupied impurity band however shows very little dispersion and it is separated 
from the valence band by about 0.08 eV. Such an impurity band is formed mainly by As-4$p$ and Mo-4$d$ orbitals.

To summarize our results for S substituting impurities, we find that most dopants create localized, spin-polarized, gap states. When increasing the dopant atomic number the spin-splitting reduces and the states move towards the VBM/CBM. Among the 
many dopants investigated As appears to give the most promising DOS for p-type doping, since the associated diamagnetic 
impurity band is very close to the VBM. In the case of n-type doping, all substituent dopants have donor states located rather far 
from the MoS$_2$ CBM and therefore they are rather localized. In addition to these considerations a look at Table~\ref{formation_0},
where we present the formation energy of the various dopants, reveals that all the S substituents possess a rather large formation 
energy, i.e. they are unlikely to form under equilibrium thermodynamical conditions. The only viable formation channel is offered by 
filling S vacancies, which have been recently demonstrated to form with relatively ease~\cite{PRL_2012_109_035503}. Therefore, it 
appears that a strategy for doping at the S site may be that of growing S poor samples and then of filling the vacancies with an 
appropriate donor/acceptor sulfur replacement. 

\subsubsection*{Substitution at the Mo site}
Next we consider substitutional doping at the Mo site with different transition metal atoms having a growing number of electrons in the 
$d$ shell. Specifically we consider all the 4$d$ elements in the periodic table ranging from Y to Cd with the only exception of radioactive 
Tc, which is replaced by Re (5$d$). In general we find that for all the substitutional atoms having a $d$ occupancy larger than Mo, the 
ground state is spin-polarized, while Y, Zr and Nb have a diamagnetic ground state (see Table~\ref{formation_0}).

\begin{figure}
\center
\includegraphics[width=8.0cm,clip=true]{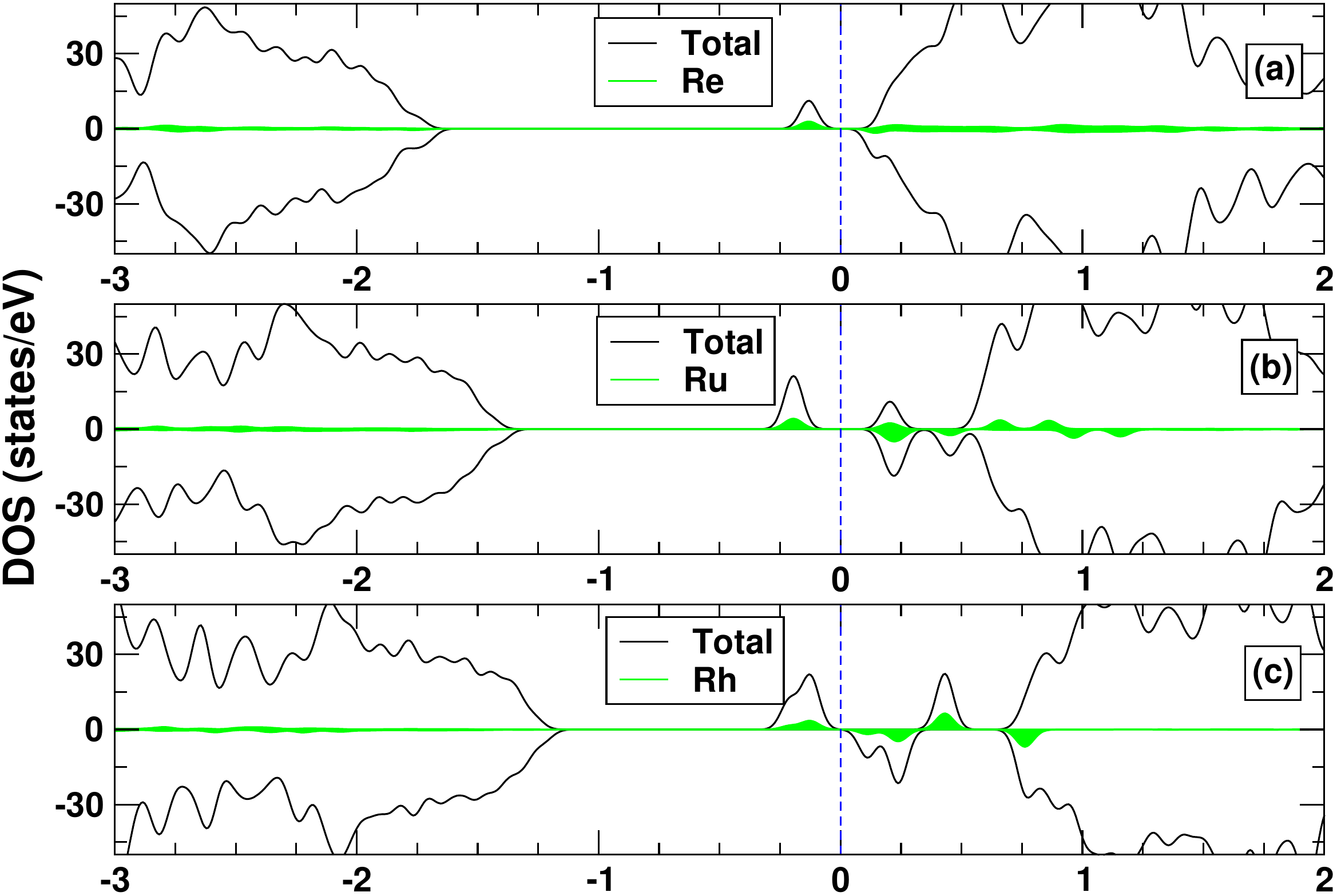}
\includegraphics[width=8.0cm,clip=true]{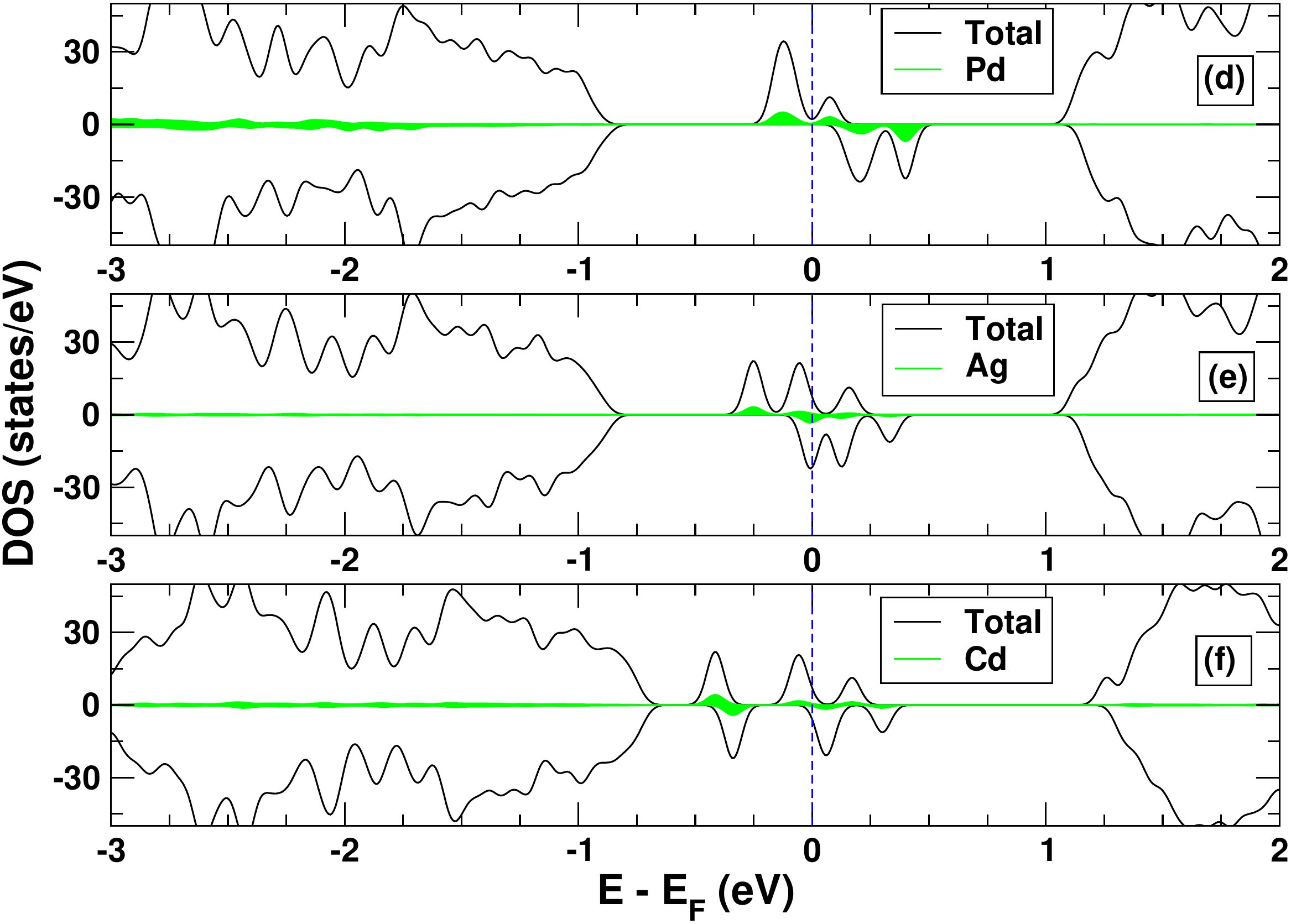}
\caption{(Color online) DOSs for a 5$\times$5 MoS$_2$ supercell in which one Mo atom is replaced by (a) Re, (b) Ru, (c) Rh, 
(d) Pd, (e) Ag and (f) Cd. The blue dashed line marks the Fermi energy, the colored shaded areas indicate the DOS projected onto the 
dopants. Note that all the DOSs are aligned to have a common Fermi level, $E_\mathrm{F}=0$.}
\label{fig:PDOS_Re}
\end{figure}

Let us again discuss the case of n-type doping first. When Re replaces Mo, one extra electron is added to the supercell. This remains 
unpaired and occupies a majority spins gap state [Fig.~\ref{fig:PDOS_Re}(a)], located 0.3~eV below the CBM, with the Fermi energy, 
$E_\mathrm{F}$ located between such state and the CBM. As a consequence the supercell has a magnetic moment of 
1~$\mathrm \mu_B$ and the DOS for this situation resembles that of Cl substituting S [Fig.~\ref{fig:PDOS_FClBrI}(b)]. If one forces a 
non-spin-polarized solution for Re-doped MoS$_2$ monolayer, the donor level is created at about 0.2~eV below the CBM. However, the 
magnetic ground state is lower in energy than the non-magnetic one by 82~meV/supercell, i.e. a diamagnetic solution is not stable. Note 
that our non-spin-polarized result is consistent to that reported for similar a calculation~\cite{ChemAsianJ_2008_3}, where a Re donor 
level at 0.19~eV below the CBM suggested that Re could be used as n-type dopant in MoS$_2$ nanotubes. Such result thus seems to 
be robust against spin polarization (the spin polarized impurity state is at 0.3~eV below the VBM). Note that single crystals Re-doped 
MoS$_2$ have been grown in the past by chemical vapor deposition~\cite{JCGrow_1999_205_543}.

When one then looks at other transition metal dopants having a $d$-orbitals occupancy larger than Re, an increasing number of 
gap states is formed. These are progressively occupied so that the excess electrons do not spill into the MoS$_2$ conduction band (see 
Fig.~\ref{fig:PDOS_Re}). As a consequence the Fermi energy always lies in the MoS$_2$ band-gap, and in fact it moves deeper into the 
band-gap as the $d$ shell filling of the dopant increases. The excess electrons first fill up the majority spin states of the host $d$-orbitals, 
until the magnetic moment of the impurity reaches the largest value of 4 $\mu_B$ for Pd. Then the magnetic moment decreases for Ag and Cd
as the additional electrons start to populate the minority spins states (see Table~\ref{formation_0}).

\begin{figure}
\center
\includegraphics[width=8.0cm,clip=true]{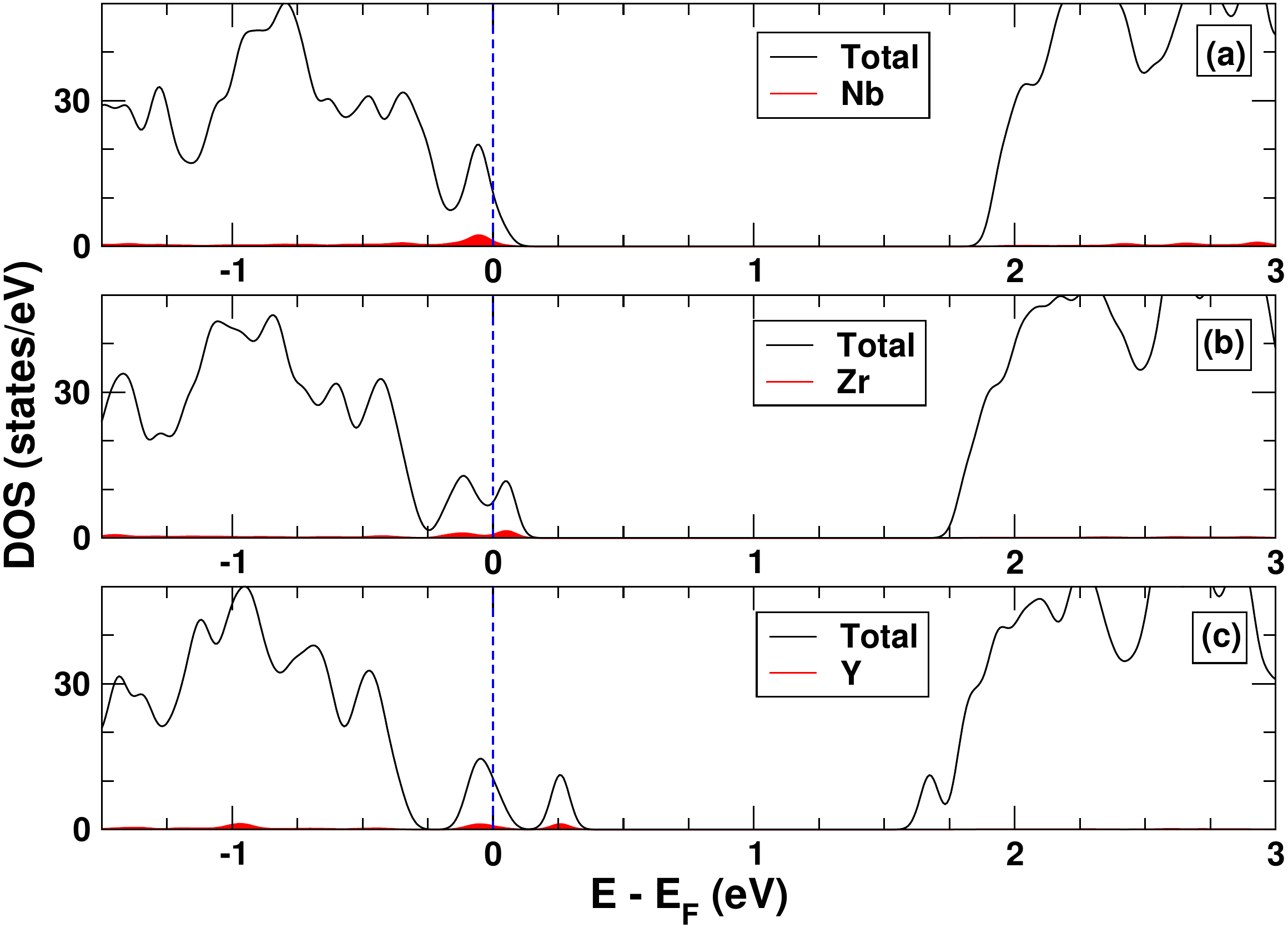}
\caption{(Color online) DOSs for a 5$\times$5 MoS$_2$ supercell in which one Mo atom is replaced by (a) Nb, (b) Zr and (c) Y. 
The blue dashed line indicates the Fermi energy, the colored shaded areas indicate the DOS projected on the dopants. Note that all 
the DOSs are aligned to have a common Fermi level, $E_\mathrm{F}=0$.}
\label{fig:PDOS_Nb}
\end{figure}

Next we move to study the possibility of obtaining p-type doping by replacing Mo with Nb, Zr and Y, which have respectively one, two
and three electrons less than Mo. The case of Nb seems to be particularly attractive. The inclusion of Nb into MoS$_2$ changes little 
the bond lengths and also the main DOS. The only notable effect is the shift of the Fermi energy below the VBM due to the one electron
removal [see Fig.~\ref{fig:PDOS_Nb}(a)]. The newly created defect states are rather delocalized and the charge excess spreads out up 
to the third nearest neighbor Mo atoms. The states around the VBM mainly originate from hybridized $d$ orbitals of Nb and Mo. The valence 
band now looks sufficiently dispersive in our DFT band-structure, and therefore the mobility is expected to be rather large. Importantly the 
formation energy for Nb-doping is -0.19~eV, in good agreement with the formation energy of -0.21~eV~\cite{PRB_2008_78_134104}, 
obtained theoretically for Nb in bulk MoS$_2$. Overall our results suggest that Nb may be a promising candidate as p-type dopant in 
MoS$_2$ monolayers. Note that an experimental study~\cite{JACS_2007_129_12549} shows that Nb-substituted (concentration range 
15-25~\%) MoS$_2$ nanoparticles can be synthesized, and that they also exhibit p-type character. Finally, for all the other two p-type 
dopants considered (Y and Zr) the ground state is also non-magnetic (see Table~\ref{formation_0}). However, in contrast to Nb doping, 
the defect state becomes less hybridized with the VBM and produces split off acceptor-levels above the VBM for both Zr and Y 
[see Fig.~\ref{fig:PDOS_Nb}(b) and (c)].

We conclude this section by evaluating the pairing energy between two Nb atoms placed in a 6$\times$6 supercell. In Table~\ref{pairing} 
we list the energy difference $\delta E$ between the configuration where the two impurities are placed at nearest neighbor positions and 
when they are placed as far as possible in the supercell. The result suggests that in case of Nb pairs the dopants tend to stay close to each 
other in the MoS$_2$ monolayer ($\delta E<$~0). Thus, based solely on the pairing energy, Nb dopants would form clusters. Clustering 
however is likely to be inhibited in by the large energy barrier for Nb diffusion in the MoS$_2$ plane.
\begin{table}
\begin{tabular}{cc}\hline
Atom-Atom & $\delta E$ (meV)\\
\hline\hline Nb-Nb&-157\\
Cs-Cs& 166\\
Nb-Cs&-218\\
\hline
\end{tabular}
\caption{Pairing energy for two impurities doping a MoS$_2$ monolayer. Calculations are based on total energy difference between
a nearest neighbor and a separated geometry in a 6$\times$6 supercell. Negative $\delta E$ indicate a tendency to clustering.}
\label{pairing}
\end{table}

Summarizing the situation for transition metal doping at the Mo site we find that, when the dopant has more $d$ electrons than Mo,
donor states are created deep inside the MoS$_2$ band-gap (at least 0.2~eV below the CBM), with Re being the dopant with the smallest 
activation energy. However, in all cases the formation energies are large (Table~\ref{formation_0}). On the contrary, p-type doping obtained
by substituting Mo with transition metals such as Nb and Zr creates acceptor states just at the VBM, and the formation energies are also 
small (Table~\ref{formation_0}). In fact, for both Nb and Zr, $E_\mathrm{form}$ is negative, indicating that substitutional doping will 
form spontaneously in MoS$_2$ monolayers (note that this is relative to the bulk reference for the dopant). Intriguingly, also
for the case of transition metal doping at the Mo site the possibility of filling vacancies remains open. In fact a recent 
experiment~\cite{APL_2012_101_102103} has provided evidence for the formation of Mo vacancies in bulk MoS$_2$ via proton irradiation.  

\subsection{Doping by adsorption}\label{absorption}
The results of the previous section show that neither Mo substitution with transition metals nor S substitution with non-metal elements 
appear as promising strategies for obtaining shallow donor states (p-doping is much more promising with Nb). Another possible 
route for obtaining n-type MoS$_2$ monolayers is by adsorbing H or alkali metals such as Li, K and Cs. This possibility is 
explored here. Then, in the second part of this section we consider adsorbed molecular ions as potential dopants with both n- and p-type 
character.
\begin{figure}
\center
\includegraphics[width=8.0cm,clip=true]{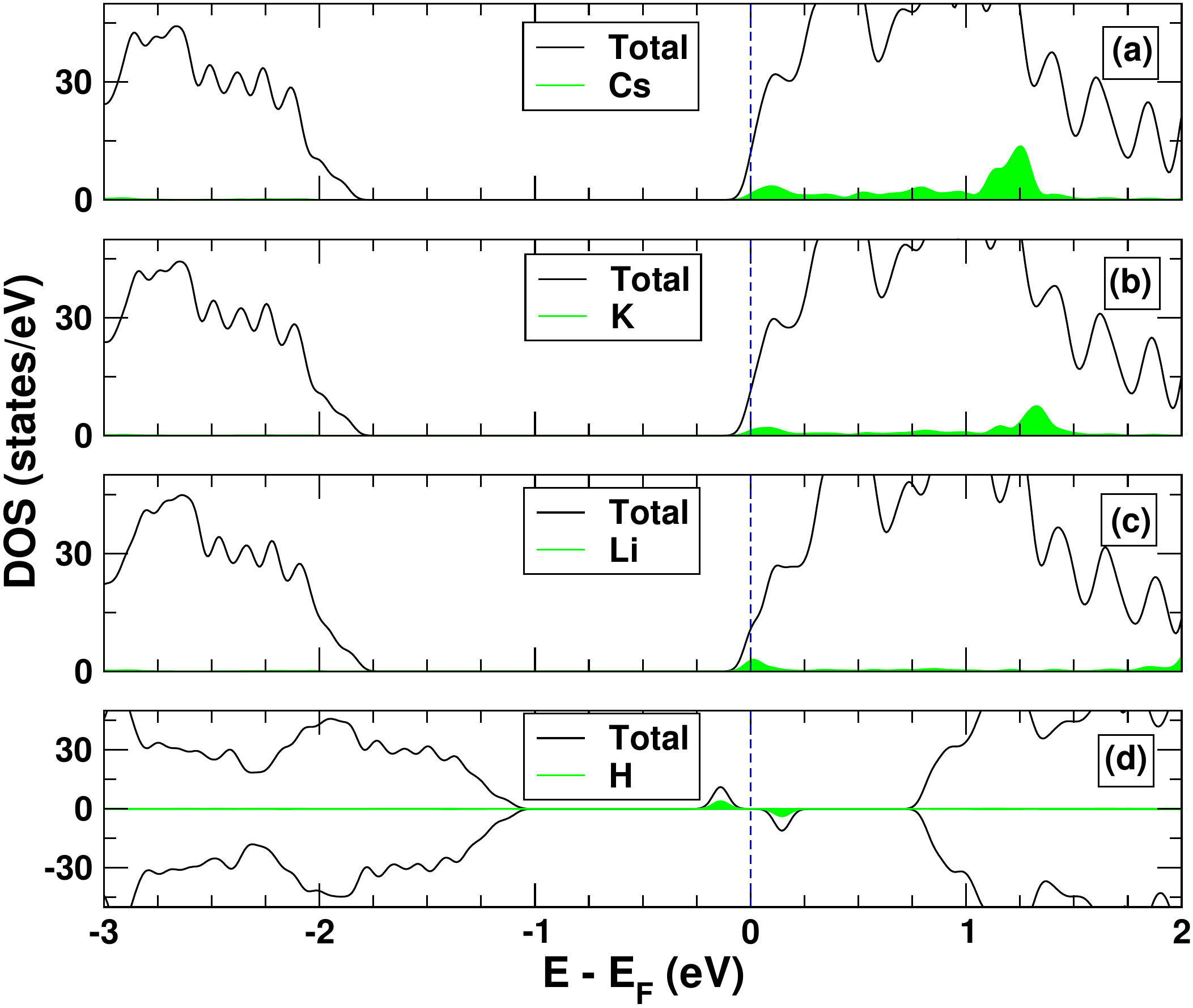}
\caption{(Color online) DOSs for a 5$\times$5 MoS$_2$ supercell in which an atom chosen between (a) Cs, (b) K, (c) Li and (d) H is adsorbed
on the MoS$_2$ surface. The blue dashed line indicates the Fermi energy, the colored shaded areas indicate the DOS projected on the 
adsorbates. Note that all the DOSs are aligned to have a common Fermi level, $E_\mathrm{F}=0$.}
\label{fig:PDOS_absorb_atom}
\end{figure}

\subsubsection*{Alkali atom adsorption}
Among all the possible adsorption sites for alkali metals on the MoS$_2$ surface, we find that the T$_\mathrm{Mo}$ one 
[see Fig. \ref{fig:ImpurityCell}(b)] is the most energetically favorable. This is also suggested experimentally~\cite{JCP_1999_111} as well 
as predicted by a previous theoretical calculations~\cite{TopCat_2002_18}. Most importantly all the adsorption energies are found 
large and negative (see Table~\ref{formation}) indicating thermodynamical stability.
\begin{table}
\begin{tabular}{c c}\hline
Adatom & E$_\mathrm{ads}$ (eV)\\
\hline \hline Cs&-0.79\\
K &-0.82 \\
Li&-0.98 \\
\hline
\end{tabular}
\caption{Theoretically calculated adsorption energy for different alkali metal adsorbed on the MoS$_2$ monolayer.}
\label{formation}
\end{table}

We start by considering Cs adsorption. When Cs is adsorbed on a MoS$_2$ monolayer there is no significant change in the geometry,
i.e. the ion binds without distorting the lattice of the host. The DOS presented in Fig.~\ref{fig:PDOS_absorb_atom}(a) shows that the 
available Cs 6$s$ electron is transferred to the CBM of the MoS$_2$ monolayer, occupying the Mo-4$d$ orbitals. In this case there
is no split off at the bottom of the conduction band and the DOS projected over the Cs 6$s$ orbitals appears uniformly spread over a
1.5~eV energy window. This essentially means that Cs adsorbed over a MoS$_2$ monolayer acts as perfect donor. A very similar
situation is found for both K and Li, with only minor quantitative differences in the spread of the DOS associated with the $s$ shell
of the dopant [see Fig.\ref{fig:PDOS_absorb_atom}(b) and Fig.\ref{fig:PDOS_absorb_atom}(c)]. In contrast, the adsorption of H produces 
a spin-split state 1~eV below the CBM, i.e. roughly at midgap of the LSDA band-gap [see Fig.\ref{fig:PDOS_absorb_atom}]. As such we
conclude that H is not a shallow donor for MoS$_2$, in good agreement with the recent theoretical results~\cite{PRL_2012_109_035503}.
\begin{figure}
\center
\includegraphics[width=8.0cm,clip=true]{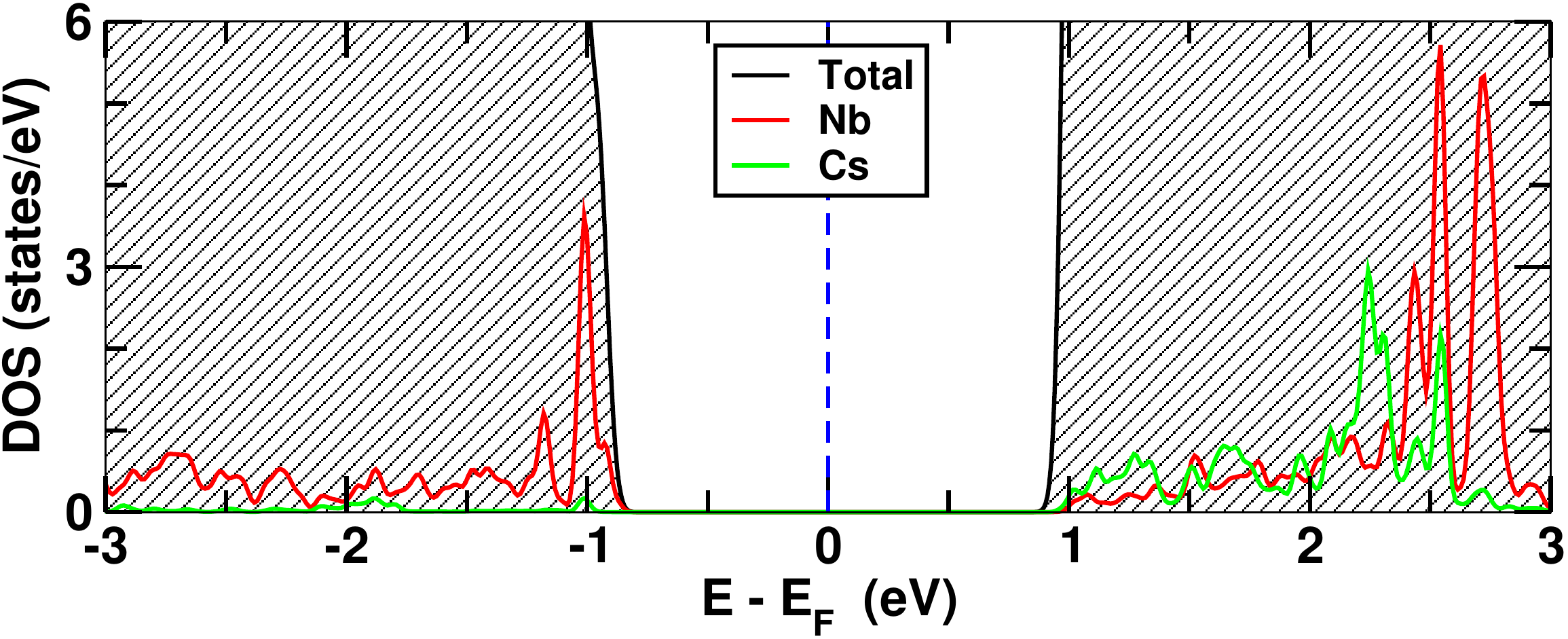}
\caption{(Color online) DOS for a MoS$_2$ supercell, where a single Mo is replaced by Nb, and where a Cs atom is adsorbed. The blue dashed 
line indicates the Fermi energy, $E_\mathrm{F}=0$, the colored curves indicate the DOS projected on the dopants.}
\label{fig:PDOS_Nb_Cs}
\end{figure}

The pairing energies between two Cs atoms and between Nb and Cs are given in Table~\ref{pairing}. Our total energy calculations suggest that pairing between Cs atoms is not energetically favorable. On the other hand Cs atoms can reduce their energy if they are adsorbed close to a Nb atom rather than on top of a Mo atom. This can be easily explained with the
electrostatic attraction between the two oppositely charged defects. 
This case of co-doping with Cs and Nb is particularly interesting, since it can be used as a strategy for
fabricating p-n junctions, or more generally devices that require both n-type and p-type conductivity. Importantly the DOS of a supercell where a Mo-substitutional Nb impurity
is bonded to a Cs adatom appears essentially identical to that of an undoped MoS$_2$ monolayer (see Fig.~\ref{fig:PDOS_Nb_Cs}), indicating 
that charge compensation is very effective. As a consequence our results suggest that an effective strategy for creating p-n hetero-junctions
must prevent the formation of Cs-Nb bonding by suppressing the migration of Cs on the surface.

In summary we find that adsorbed alkali metals release their valance $s$ electron to the MoS$_2$ conduction band, and that the adsorption 
energies are large and negative, meaning that the doping will occur easily. Therefore, adsorbed alkali metals, with the exception of H, appear as ideal 
candidates for doping n-type MoS$_2$. To put this result in context we remark that various experimental studies have demonstrated the possibility of 
dope n-type bulk MoS$_2$ by intercalating alkali metals \cite{Adv_1987_36}. Here we demonstrate that the same is also possible at 
the single layer level.

\subsubsection*{Adsorption of molecular ions}

\begin{figure}
\center
\includegraphics[width=8.0cm,clip=true]{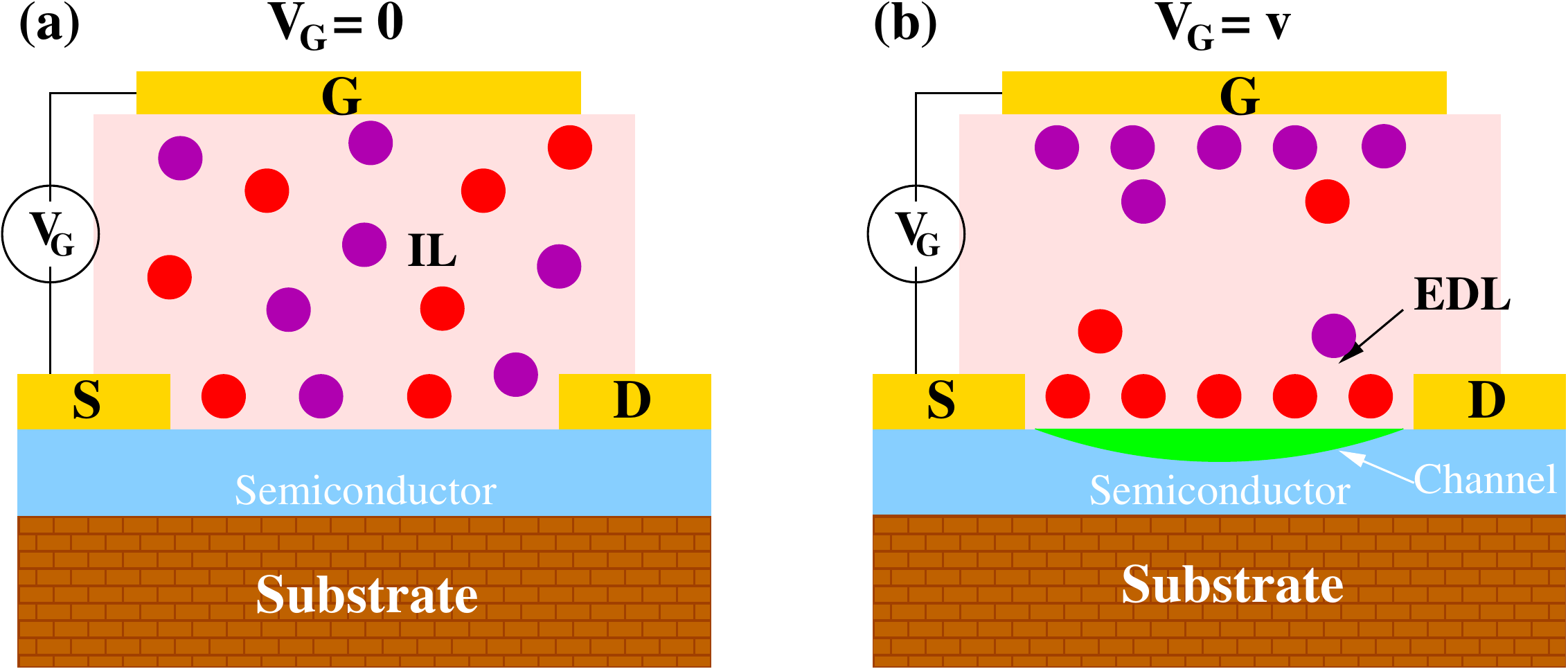}
\caption{Schematic diagrams of the electric double layer transistor operation with an ionic liquid electrolyte. Here, ``S'' indicates the source, 
``D'' indicates the drain, and ``G'' indicates the gate. The solid circles denote the ions in the liquid, where the different colors represent cations 
and anions. (a) When no gate voltage is applied (V$_\mathrm{G}$=0), cations and anions are uniformly distributed and both adsorbed at the 
interface of the semiconductor with equal probability. (b) When a finite gate bias is present (V$_\mathrm{G}=v$), cations or anions adsorb 
predominantly at the gate electrode, depending on the bias polarity, and the oppositely charged ions adsorb predominantly on the semiconductor. 
These adsorbed ions lead to an accumulation of a screening charge at the semiconductor surface, which implies a large surface carrier 
concentration.}
\label{fig:schematic_ion}
\end{figure}

Electric double layer (EDL) transistors with ionic liquids (ILs) employed as gate dielectrics have recently emerged as promising devices, 
where the electrical properties of a solid can be controlled by electrostatic carrier doping \cite{NL_2012_12_2988, APL_2011_98_012102}. 
A schematic diagram for such a device is shown in Fig.~\ref{fig:schematic_ion}. By applying a gate bias to the ionic liquid, an electric double 
layer is formed between the interface of the liquid and the solid (see Fig.~\ref{fig:schematic_ion}). Here the polarity of the molecular ions on 
the surface of the semiconductor can be reversed by changing the polarity of the gate bias. The accumulation of high charge carrier densities 
is possible ($\sim$10$^{14}$ cm$^{-2}$) and this can be modulated by the applied gate voltage. In some case the accumulation can be so
large to lead to phase transitions in the solid. Metal-superconductor and metal-insulator transitions have been demonstrated in KTiO$_3$ 
\cite{NatureNano_2011_6_408} and VO$_2$ \cite {NL_2012_12_2988}, respectively. The operation of highly flexible MoS$_2$ thin-film 
transistors in an ionic liquid gated dielectric has been demonstrated recently \cite{NL_2012_12_4013}. Here we discuss the effects of the 
adsorbed molecular ions on the MoS$_2$ monolayer.
\begin{figure}
\center
\includegraphics[width=8.0cm,clip=true]{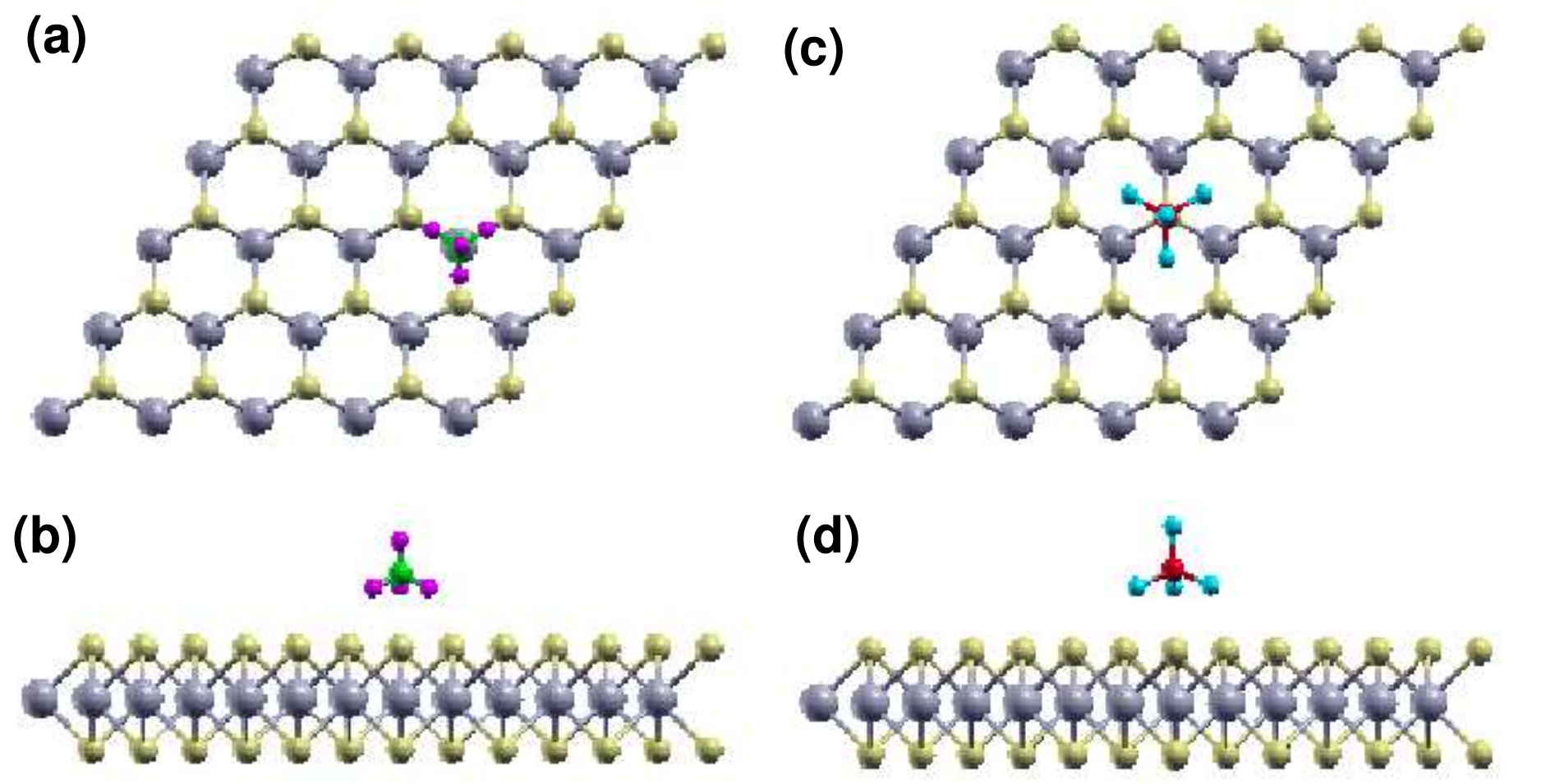}
\caption{(Color online) The optimized geometries for (a, b) NH$_4^+$ and (c, d) BF$_4^-$ ions adsorbed on a MoS$_2$ monolayer. 
In (a) and (c) the top view is shown, whereas the side view is in (c) and (d). Pink spheres indicate H, green spheres N, cyan spheres F 
and red spheres B.}
\label{fig:adsorb_ion}
\end{figure}
\begin{figure}
\center
\includegraphics[width=8.0cm,clip=true]{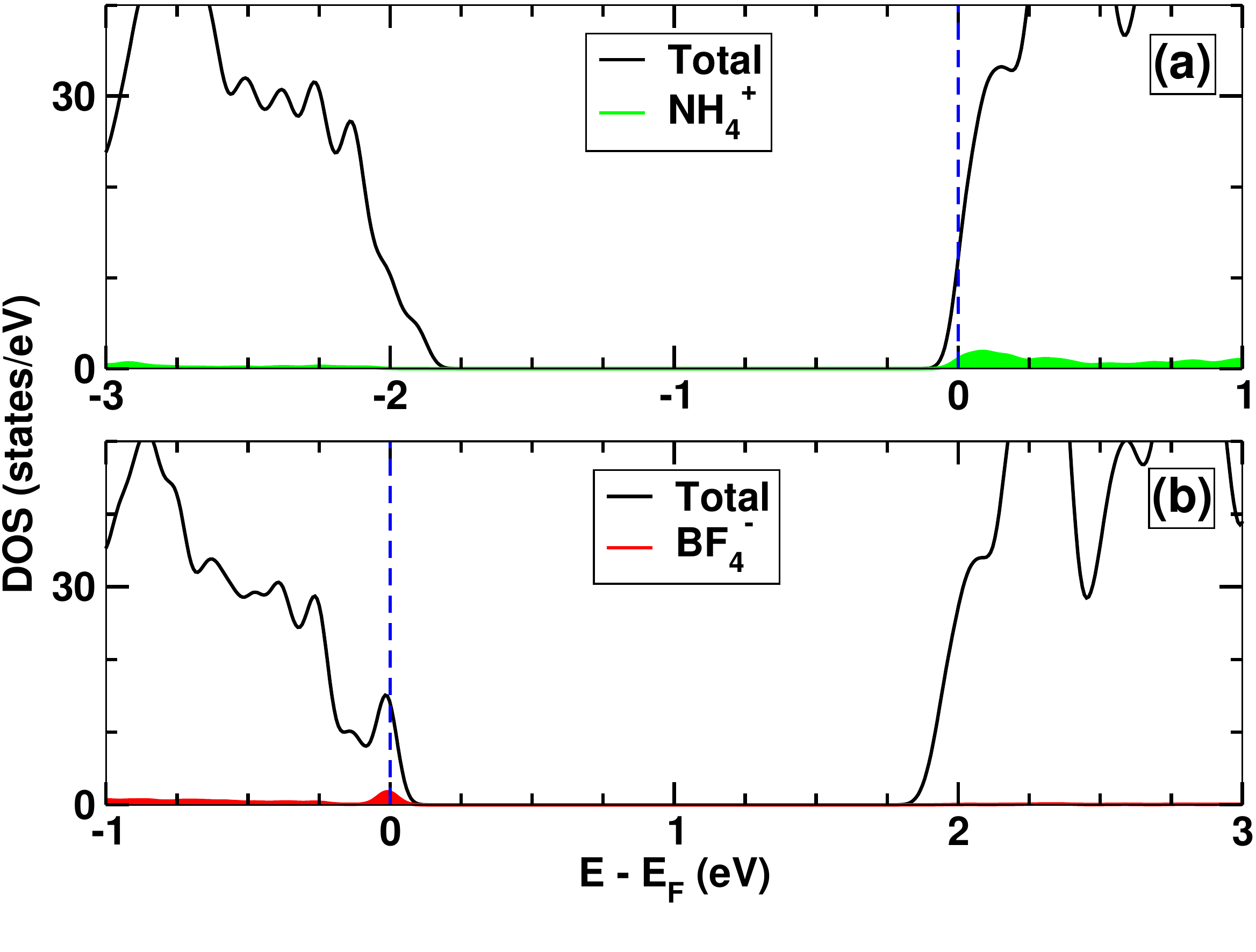}
\caption{(Color online) DOS for (a) NH$_4^+$ and (b) BF$_4^-$ molecule adsorbed on MoS$_2$. The results reported here are for a 5$\times$5 
supercell. The blue dashed line indicates the Fermi energy, the colored shaded areas indicate the DOS projected over the adsorbates. Note that 
the DOSs have the Fermi level at 0 ($E_\mathrm{F}=0$) but they are aligned according to their valence and conduction bands.}
\label{fig:PDOS_absorb_ion}
\end{figure}

A variety of different ionic liquids are experimentally available. Here we consider NH$_4^+$ as cation and BF$_4^-$ as anion, and we place them
on the $5\times5$ MoS$_2$ supercell (see Fig. \ref{fig:adsorb_ion}). Such geometry corresponds to an impurity density of $\sim$10$^{14}$ cm$^{-2}$. 
We find that the most energetically favorable adsorption site is T$_\mathrm{S}$ for the cation and T$_\mathrm{Mo}$ for the anion. In 
Fig.~\ref{fig:PDOS_absorb_ion}(a) the DOS is shown for NH$_4$ adsorption. It can be seen that the NH$_4$ releases one electron into the MoS$_2$ 
conduction band, resulting in a n-type MoS$_2$ monolayer channel. In contrast, when BF$_4$ is adsorbed, one electron is transferred from MoS$_2$ 
to the molecule [see Fig.~\ref{fig:PDOS_absorb_ion}(b)]. As a consequence the Fermi energy shifts below the MoS$_2$ VBM, so that the system becomes 
p-type. This also confirms the fact that the adsorbed molecules are indeed ionized to NH$_4^+$ and BF$_4^-$ when put on MoS$_2$. In a liquid 
gated transistor arrangement, the anions/cations are brought close to the surface of the channel by applying a bias at the gate. Therefore, for this 
system, depending on the polarity of the gate bias the MoS$_2$ is indeed predicted to have a n- or p-type conducting channel. Clearly this effect is 
maximized for a single layer of MoS$_2$ when compared to thicker structures. In fact, conductivity measurements~\cite{NL_2012_12_1136} on 
MoS$_2$ show that EDL gating becomes less efficient when going from a thin layer to bulk. Moreover, depending on the polarity of the bias, it is found 
that the conductivity indeed switches from n-type to p-type.

\subsection{Robustness of the results against the choice of XC functional:HSE06}

Finally, in order to verify that the impurity level alignment presented in the previous sections is robust against the level of approximation taken
for the DFT exchange-correlation energy, we here repeat our calculations by using the HSE06 functional. The MoS$_2$ lattice constant obtained 
with HSE06 is 3.155~\AA, which is only slightly larger than the LSDA value of 3.137~\AA. For a pristine MoS$_2$ monolayer at this relaxed lattice 
constant HSE06 exhibits a direct band-gap of 2.2~eV (compared to a LDA band-gap of 1.86~eV). This is rather similar to the band-gap obtained 
by applying the atomic self-interaction correction scheme~\cite{arXiv:1301.2491}.
\begin{figure}[h]
\center
\includegraphics[width=8.0cm,clip=true]{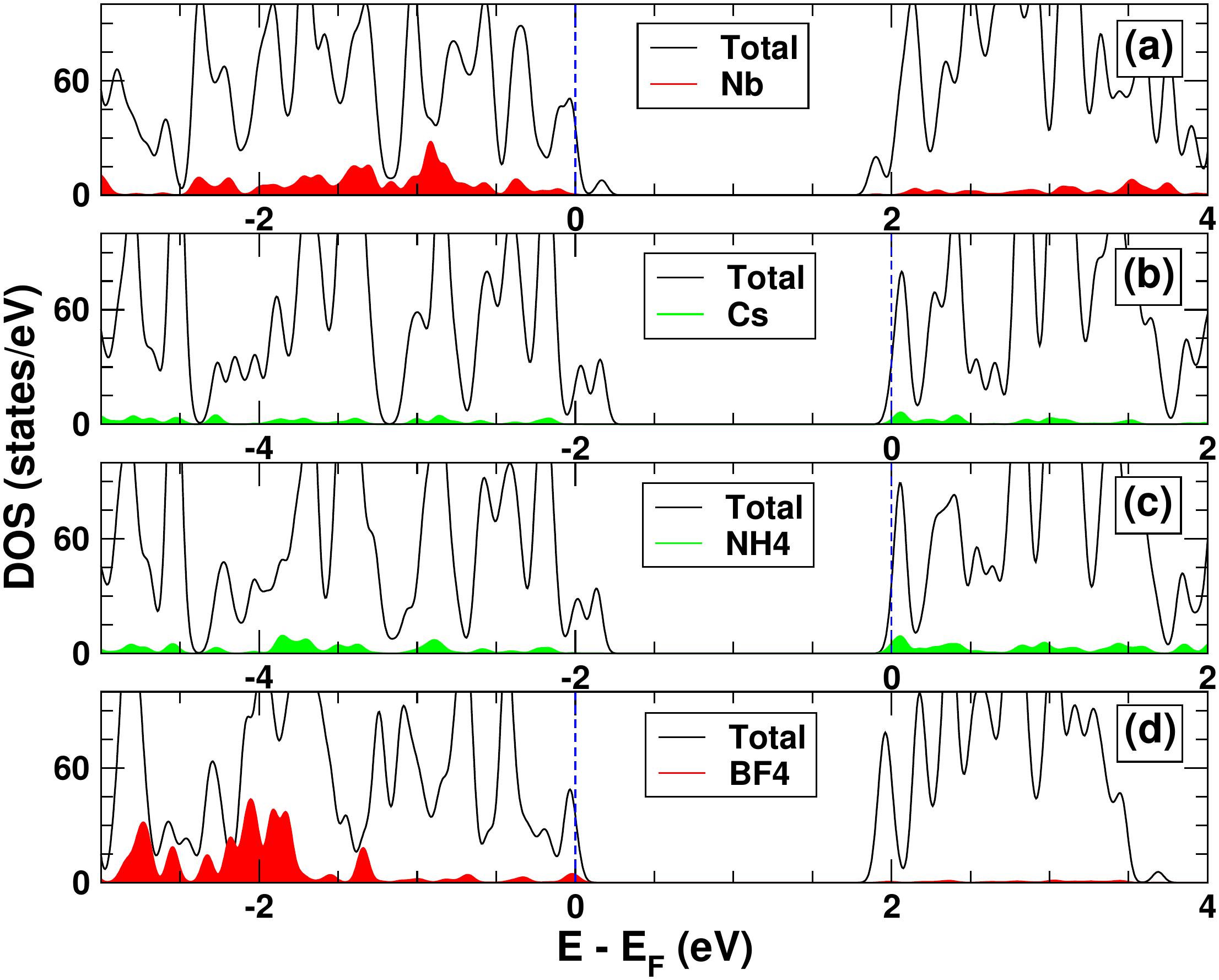}
\caption{(Color online) DOSs for the doped monolayer MoS$_2$, calculated with the HSE06 functional. We report results for (a) Nb substitutional 
at the Mo site, and for (b) Cs, (c) NH$_4^+$ and (d) BF$_4^-$ adsorbed on the MoS$_2$ surface. The blue dashed line indicates the Fermi energy 
($E_\mathrm{F}=0$), the colored shaded areas indicate the DOS projected on the adsorbates. In the plot we have aligned all the conduction and 
valence bands.}
\label{fig:PDOS_HSE}
\end{figure}

In Fig.~\ref{fig:PDOS_HSE} we report our results for four representative doping strategies, namely for Nb substitutional doping, as well as for Cs, 
NH$_4$ and BF$_4$ adsorption. These correspond to the most promising dopants for either n-type or p-type doping as obtained at the LSDA level. 
The calculations are carried out with the same structures used in the previous sections, in order to make a direct comparison between results 
obtained with the two functionals used. We note, however, that the MoS$_2$ band-gap is very sensitive to changes in the lattice 
constant~\cite{PRB_2012_85_033305}, so that the HSE06 band-gap at the LSDA lattice constant of 3.137~\AA~ becomes 2.31~eV. We 
find that for all the four cases the main difference between the LSDA and the HSE06 results is just a change in the band-gap, while the impurity levels 
with respect to the relative Fermi energy are placed at the same position. Therefore we can conclude that the four most promising candidates
for doping MoS$_2$ monolayers remain those identified by our initial LSDA analysis.

\section{Conclusion}
We have carried out a systematic study on the changes to the electronic structure of a MoS$_2$ monolayer induced by various dopants. 
This study has the aim of identifying possible doping strategies for MoS$_2$ monolayers to be used in devices manufacturing. We have 
considered first substitutional doping at both the Mo and the S site, as well as doping by adsorption. In general, S substitution with non-metals 
and Mo substitution with transition metals create deep donor levels inside the band-gap of the MoS$_2$ monolayer for most of the dopants
considered. However, with substitutional doping we find that it is possible to obtain p-type MoS$_2$ by replacing a Mo atom with Nb. 
In contrast n-type doping does not appear possible since we have found deep donor levels for all the substitutions investigated (either at 
the Mo or the S site). More promising n-type doping can be achieved by adsorbing alkali metals on the surface of MoS$_2$. Finally, as a 
last class of dopants, we have considered the adsorption of ionic molecules, which occurs during the electric double layer formation in 
MoS$_2$ embedded in an ionic liquid. These show high potential for inducing large carrier concentrations within the MoS$_2$ monolayer, 
in the form of both electrons or holes. 

\section*{Acknowledgments}

This work is supported by Science Foundation of Ireland (Grant No. 07/IN.1/I945) and by CRANN. IR acknowledges financial support from 
the King Abdullah University of Science and Technology ({\sc acrab} project). We thank Trinity Centre for High Performance Computing 
(TCHPC) for the computational resources provided. 


\end{document}